\documentclass[jkps,preprint,fleqn,showpacs,showkeys]{revtex4}
\usepackage{amssymb}
\usepackage{amsmath}
\usepackage{bm}

\newcommand{\Fs}{\,^*\! F} 
 \newcommand{\bJ}{\bm{J}}

\newcommand{\bbeta}{\bm{\beta}} 
\newcommand{\bomega}{\bm{\omega}} 
\newcommand{\bmath}[1]{\bm{#1}}
 \newcommand{\bn}{\bm{n}}
\newcommand{\bk}{\bm{k}} 
\newcommand{\bB}{\bm{B}} \newcommand{\bE}{\bm{E}}
\newcommand{\bH}{\bm{H}} \newcommand{\bD}{\bm{D}}
\newcommand{\bL}{\bm{L}} 
 \newcommand{\bS}{\bm{S}}

\newcommand{\ntext}[1]{\quad\mbox{#1}\quad}
\newcommand{\spr}[2]{\bm{#1} \!\cdot\! \bm{#2}}
\newcommand{\vpr}[2]{\bm{#1} \!\times\! \bm{#2}}
\newcommand{\vdiv}[1]{\spr{\nabla}{#1}}

\newcommand{\vcurl}[1]{\vpr{\nabla}{#1}}

\newcommand{\Pd}[1]{\partial_{#1}} 
 \newcommand{\ortt}[1]{
\bm{i}_{\hat{#1}} } 
 \newcommand{\upin}[1]{{^{\hat{#1}}}}

\begin{document}
\setcounter{page}{0}
\title{Blandford-Znajek mechanism versus Penrose process}
\author{ Serguei S.\surname{Komissarov}}
\email{serguei@maths.leeds.ac.uk}
\affiliation{Department of Applied Mathematics, 
The University of Leeds, Leeds, LS2 9GT}
\date{Received/Accepted}

\begin{abstract}  
During the three decades since its theoretical discovery the Blandford-Znajek
process of extracting the rotational energy of black holes has become 
one of the foundation stones in the building of modern relativistic 
astrophysics. However, it is also true that for a long time its physics 
was not well understood, as evidenced by the controversy that 
surrounded it since 1990s. Thanks to the efforts of many theorists 
during the last decade the state of affairs is gradually improving. 
In this lecture I attempt to explain the key ingredients of this process 
in more or less systematic, rigorous, and at the same time relatively simple 
fashion. A particular attention is paid to the similarities and differences between 
the Blandford-Znajek and Penrose processes. To this purpose I formulate the 
notion of energy counter flow.  The concept of horizon membrane 
is replaced with the concept of vacuum as an electromagnetically active medium.  
The effect of negative phase velocity of electromagnetic waves in the black 
hole ergosphere is also discussed. 

\end{abstract}
                                                                                          
\maketitle

\section{Introduction}
\label{introduction}
 
The Blandford-Znajek process is widely believed to be one of the most promising mechanisms  
for powering the relativistic outflows from black holes in a variety of astrophysical 
phenomena including Active Galactic Nuclei, Gamma-Ray-Bursts, and Galactic X-ray Binaries.   
However, even 30 years after the seminal paper of Blandford and Znajek\cite{BZ77} 
the nature of this mechanism is still under investigation by theorists 
\cite{BK00,Pun01,K04a,Oka,L06}. 
The origin of electromotive force that drives the electric currents in black hole 
magnetosphere is the focal point of these investigations, with emphasis being made either 
on the event horizon \cite{TPM}, the ergosphere \cite{Pun01,K04a}, or the pair creation 
surfaces \cite{BK00,Oka}. 

The Blandford-Znajek process is purely electromagnetic. Although highly 
conducting plasma is a key ingredient, its inertia and energy 
are assumed to be vanishingly small. This assumption leads to a puzzle   
as it leaves no place for a matter-dominated unipolar inductor, in contrast to 
say the stellar magnetospheres where the the role of Faraday disk is played by a rotating 
magnetised star.  One way to ``solve'' this puzzle is to endow the event horizon with 
the properties of rotating conductor (among many other properties) \cite{TPM}. 
However, this construction is obviously artificial, which has been reflected in its name --
``Membrane paradigm''. Its shakiness has revealed itself when 
Punsly and Coroniti formulated the so-called ``causality paradox'' of Blandford-Znajek 
process \cite{PCa,PCb,Pun01} -- basically the event horizon cannot serve as a unipolar inductor 
because it is causally disconnected from the black hole exterior. 

Alternative magnetohydrodynamic (MHD) mechanisms have been put forward which operate 
on the principles similar to the mechanical process of Penrose\cite{Penrose}. In these
mechanisms \cite{PCb,Pun01,Koi02} it was proposed that 
(i) The unavoidable  rotation of plasma inside the ergosphere forces 
the magnetic field lines frozen in this plasma to rotate as well. The magnetic twist 
propagates away from the black hole, resulting in an outgoing Poynting flux;  
(ii) As a feedback action, the magnetic forces push plasma into orbits with negative 
mechanical energy at infinity before they plunge into the black hole, resulting in 
an outgoing flux of mechanical energy-at-infinity. The total energy 
flux is conserved but it changes its nature from almost purely mechanical close to 
the horizon to almost purely electromagnetic far away from it. In a way, the ergospheric 
plasma and the magnetic field play the roles similar to those of the negative and positive 
energy particles in the Penrose process.

Recently, a number of attempts have been 
made to clarify the issue via time-dependent numerical simulations, both electrodynamic 
\cite{K01,K04a,McK06} and magnetohydrodynamic \cite{K04b,Koi04,K05}. Their results have shown 
that (i) the Blandford-Znajek mechanism does not clash with causality, 
(ii) the particle inertia does not have to be a dynamical factor, 
(iii) the regions of negative mechanical energy-at-infinity do not readily develop 
(They are not seen in steady-state solutions, but may occur as transients 
in some highly variable solutions). Moreover, these conclusions have found strong support 
in recent theoretical studies too \cite{BK00,K04b,L06}. 
Thus, now there is hardly a doubt in that the Blandford-Znajek process is a fully viable 
physical process, and that the controversy that has been surrounding it for a long time was
just due to the lack of full understanding of its operation.     

Although the MHD models based on influx of negative mechanical energy-at-infinity 
do not seem to work the general idea that any process of extraction of the rotational energy 
of black holes is based on an influx of negative energy-at-infinity of one sort of another 
looks very attractive. Indeed, because the event horizon works as a one-way membrane allowing 
only ingoing waves and particles the only way to change its mass is via influx of  
matter and fields carrying negative energy-at-infinity. The mechanical Penrose process 
\cite{Penrose} and 
the superradiant scattering \cite{Sta73,SC74} are the clear examples of this sort.     
Even the Hawking radiation \cite{Hawking} from non-rotating black holes fits within this 
concept (e.g. see the neat discussion in \cite{Sch}). Therefore, the natural expectation 
is that the Blandford-Znajek process is yet another variation on this theme.   

Although the fully covariant theory of black hole electrodynamics has its indisputable merits, 
the 3+1 approach has been gaining popularity in the astrophysical community as it 
allows us to utilise both the well know tools of vector
calculus and the physical intuition developed in classical electrodynamics. The most 
popular version of this approach is based on the introduction of only two 
spatial vectors, $\bE$ and $\bB$, electric and magnetic field as measured by an observer 
at rest in the absolute space \cite{MT}. However, a different approach, originally 
suggested by Tamm\cite{Tamm} back in 1920s and later developed by other authors 
\cite{LL51,Pleb}, 
where all four vectors of Maxwell theory are introduced (including $\bH$ and $\bD$) and 
the 3+1 equations of black electrodynamics take exactly the same form as the Maxwell 
equations in electromagnetically active medium, 
has been rediscovered recently both by astrophysicists \cite{K04a} and physicists 
\cite{MLS05a,Ulf}. Such ``medium'' is present even 
in absence of plasma and one can understand it as vacuum being endowed with electromagnetic 
properties in strong gravitational field. 

A very interesting and somewhat indirectly related to our subject development has taken place 
recently in terrestrial physics. A very unusual materials, so called ``meta-materials'', 
that allow simultaneously negative values for both magnetic permeability
and electric permittivity (within a limited waveband) have been manufactured in laboratory. 
In such a medium, hypothesised in the theoretical study by Veselago\cite{Ves}  back in 
1960s, many properties of waves become rather unusual. For example, vectors $\bE$, $\bH$, and 
the wave vector $\bk$ form a left-handed triplet, the refraction index is negative, and  
the Poynting vector points in the opposite direction to $\bk$, the direction of phase 
propagation. 

Obviously, these new developments stimulated research into wave 
propagation in mediums with peculiar properties and several recent studies have 
been devoted to the propagation of electromagnetic waves in strong gravitational field 
\cite{MLS05a,MLS05b,LMS05,SML,Ulf,McCall}.     
In particular, Mackay et al.\cite{MLS05b} demonstrated that in the case of rotating 
black holes the Poynting vector can also be antiparallel to the wave vector. 
The results of their numerical study seem to indicate 
that this effect, which they called this the ``Negative Phase Velocity'' effect, can occur 
only inside the ergosphere.  Later they argued that it is different from the similar 
effect of superradiance \cite{SML}.  Again, these findings invite us to
compare the Blandford-Znajek process with the Penrose one.    

This paper is structured as follows. In Sec.2 we recall the 3+1 splitting of Kerr space-time 
in Boyer-Lindquist foliation. Sec.3 describes the 3+1 splitting of electrodynamics. In Sec.4  
we analyse the NPV effect in a somewhat different way to Mackay et al.\cite{MLS05b} 
and derive a very simple analytical condition with clear physical meaning. 
In Sec.5  we consider stationary force-free conductive magnetospheres of Kerr black holes of 
the in an attempt to put the Blandford-Znajek process on the same footing as the Penrose 
process. In Sec.6 we discuss the nature of the electromotive force of the process. 
In Sec.7 we summarise our discussion.  Throughout the paper we employ 
relativistic units where the black hole mass $M=1$ and the speed of light $c=1$. 

\section{3+1 splitting of space-time}

Following Macdonald \& Thorne\cite{MT}  we adopt the foliation
approach to the 3+1 splitting of spacetime and use the notation introduced by 
York\cite{York}.  In this approach     
the time coordinate $t$ parameterises a suitable filling of spacetime  
with space-like hyper-surfaces described by the 3-dimensional metric tensor 
$\gamma_{ij}$. These hyper-surfaces may be regarded as the 
``absolute space'' at different instances of time $t$. Below we 
describe a number of useful results for further references.     
If $\{x^i\}$ are the spatial coordinates of the absolute space then   
\begin{equation} 
  ds^2 = (\beta^2-\alpha^2) dt^2 + 2 \beta_i dx^i dt + 
         \gamma_{ij}dx^i dx^j
\label{metric} 
\end{equation}
where $\alpha$ is called the ``lapse function'' and  
$\bbeta$ is the ``shift vector''. 
The 4-velocity of the local fiducial observer, 'FIDO',
which can be described as being at rest in the absolute space, is   
\begin{equation} 
   n_\mu = (-\alpha,0,0,0).  
\label{ncov} 
\end{equation}
The spatial components of the projection tensor, which is used to 
construct pure spatial tensors, 
\begin{equation} 
   \gamma_{\alpha\beta} = g_{\alpha\beta} + n_\alpha n_\beta ,
\end{equation}
coincide with the components of the spatial metric $\gamma_{ij}$.  
Other useful results are 

\begin{equation} 
   n^\mu = \frac{1}{\alpha}(1,-\beta^i),
\label{ncon} 
\end{equation}
\begin{equation} 
   g^{t \mu} = -\frac{1}{\alpha} n^\mu,  
\label{gcov} 
\end{equation}
\begin{equation} 
   g = -\alpha^2 \gamma,    
\label{dets} 
\end{equation}

\noindent
where 
$$
 \beta^i=\gamma^{ij}\beta_j  , \quad  
 g = \det{g_{\mu\nu}}, \quad 
 \gamma = \det{\gamma_{ij}}.  
$$
$-\beta^i$ are the velocity components of local FIDO relative to the spatial 
grid as measured in the coordinate basis of spacial vectors, $\{ \Pd{i}\}$, 
and derived using the coordinate time $t$ \cite{MT}.

In this paper we will use the Boyer-Lindquist coordinates $\{t,\phi,r,\theta\}$.
In these coordinate the Kerr metric reads  
\begin{equation}
   ds^2=g_{tt}dt^2+2g_{t\phi}dtd\phi+g_{\phi\phi}d\phi^2 +
        g_{rr}dr^2 + g_{\theta\theta}d\theta^2,
\label{BLM}
\end{equation}
where
\begin{equation}
\nonumber
\begin{array}{lll}
g_{tt} &=& (2r/\rho^2)-1,\\
g_{t\phi} &=& -2ra\sin^2\!\theta/\rho^2,\\
g_{\phi\phi} &=& \Sigma\sin^2\!\theta/\rho^2,\\
g_{rr} &=& \rho^2/\Delta,\\
g_{\theta\theta} &=& \rho^2,
\end{array}
\end{equation}
and
\begin{equation}
\nonumber
\begin{array}{lll}
  \rho^2 &=& r^2 +a^2\cos^2\!\theta, \\
   \Sigma &=& (r^2+a^2)^2-a^2\Delta\sin^2\!\theta, \\
   \Delta &=& r^2+a^2-2r.
\end{array}
\end{equation}
Here $a =J/M^2$, where $J$ is the angular momentum of the black hole 
and $M$ is its mass, is the parameter that measures the rotation rate of 
the black hole. In the relativistic units adopted in this paper 
$-1<a<+1$, but we will consider, without any loss of generality, 
only anti-clockwise rotation for which $a$ is positive. 
Simple calculations show that 
\begin{equation}
\begin{array}{lll}
    g &=& -\rho^4\sin^2\theta,\\
    \beta^2 &=& 4a^2r^2\sin^2\theta/\Sigma,\\
    \alpha^2 &=& \Delta\rho^2/\Sigma,\\
    \gamma &=& \Sigma\sin^2\theta\rho^2/\Delta.
\end{array}
\end{equation}
The Boyer Lindquist FIDO is in the state of purely azimuthal
motion with constant angular velocity,    
\begin{equation}
    \bbeta = (\beta^\phi,0,0), \quad
    \beta^\phi = -2ar/\Sigma. 
\label{BLFIDO}
\end{equation}
From (\ref{ncov}) one finds that
$n_\phi=0$ and, thus, FIDO is also a zero angular momentum observer 
(ZAMO, Bardeen at al.,1973.) 

The condition $\Delta=0$ defines the two event horizons of Kerr black hole,
\begin{equation}
  r=r_\pm = 1 \pm\sqrt{1-a^2}. 
\end{equation}
Only the outer horizon, $r=r_+$ is of interest in the current context. 
The condition 
\begin{equation}
g_{tt}=0 \ntext{or} \alpha^2=\beta^2
\end{equation}
defines the ergosphere, the surface 
inside which even light is dragged into rotation in the same sense as the black hole. 
Finally, we notice that the metric coefficients of Boyer-Lindquist coordinates do not 
depend on $t$ and $\phi$ which leads to conservation of energy and angular momentum.

\section{3+1 equations of black hole electrodynamics}

The covariant Maxwell equations are (e.g. Jackson(1975)):
\begin{equation}
  \nabla_\beta  \Fs^{\alpha \beta} = 0,
\label{Maxw1}
\end{equation}
\begin{equation}
  \nabla_\beta  F^{\alpha \beta} =  I^\alpha,
\label{Maxw2}
\end{equation}
where $F^{\alpha\beta}$ is the Maxwell tensor of the electromagnetic
field, $\Fs^{\alpha\beta}$ is the Faraday tensor and
$I^\alpha$ is the 4-vector of the electric current.
In the following we assume that
\begin{equation} 
 \Pd{t}\alpha = \Pd{t}\beta_i = \Pd{t} \gamma_{ij} =0,        
\label{stat_metr} 
\end{equation}
that implies a stationary space-time and suitable selections of its foliation 
and spatial coordinates. Obviously, this applies to the metric of Kerr space-time 
in the Boyer-Lindquist coordinates.  

Now we introduce the spatial vectors $\bB$, $\bE$, $\bH$, $\bD$, 
$\bJ$ and the spatial scalar $\rho$ via
\begin{equation}
     B^i=\alpha \Fs^{it}
\label{B1}
\end{equation}   
\begin{equation}
   E_i =\frac{\alpha}{2} e_{ijk} \Fs^{jk},
\label{E1}
\end{equation}
\begin{equation}
     D^i=\alpha F^{ti},
\label{D1}
\end{equation}   
\begin{equation}
    H_i =\frac{\alpha}{2} e_{ijk} F^{jk}.
\label{H1}
\end{equation}  
\begin{equation}
    \rho=\alpha I^t, \quad J^k=\alpha I^k.
\label{rhoJ}
\end{equation}
where $e_{ijk} = \sqrt{\gamma} \epsilon_{ijk} $
is the Levi-Civita pseudo-tensor of the absolute space. In terms of 
these vectors the 3+1 splitting of equations (\ref{Maxw1},\ref{Maxw2}) 
reads 
\begin{equation}
   \vdiv{B}=0,
\label{divB}
\end{equation}
\begin{equation}
   \Pd{t}\bB + \vcurl{E} = 0,
\label{Faraday}
\end{equation}
\begin{equation}
\label{divD}
   \vdiv{D}=\rho,
\end{equation}
\begin{equation}
   -\Pd{t}\bD + \vcurl{H} = \bJ
\label{Ampere}
\end{equation}  
\noindent
where $\nabla$ is the covariant derivative of the absolute space \cite{K04a}.

In vacuum  (as well as in highly ionised plasma) the electric and magnetic 
susceptibilities vanish and the Faraday tensor is simply dual to the Maxwell 
tensor
\begin{equation}
\Fs^{\alpha  \beta} = \frac{1}{2} e^{\alpha \beta \mu \nu} F_{\mu \nu},
\qquad
F^{\alpha  \beta} = -\frac{1}{2} e^{\alpha \beta \mu \nu} \Fs_{\mu \nu},
\label{F}
\end{equation}
where
\begin{equation}
e_{\alpha \beta \mu \nu} = \sqrt{-g}\,\epsilon_{\alpha \beta \mu \nu},
\label{LCt}
\end{equation}
is the Levi-Civita alternating tensor of spacetime
and  $\epsilon_{\alpha \beta \mu \nu}$ is the four-dimensional Levi-Civita
symbol. This gives us to find the following constitutive equations relating 
$E,D,H$, and $B$:
\begin{equation}
     \bE = \alpha \bD + \vpr{\bbeta}{B},
\label{E3}
\end{equation}
\begin{equation}
     \bH = \alpha \bB - \vpr{\bbeta}{D}.
\label{H3}
\end{equation}
These equations allow us to conclude that {\it vacuum of curved space-time behaves as 
an instantly-responding bi-anisotropic medium}. 

The covariant energy-momentum equation for electromagnetic field is      
\begin{equation}
     \nabla_\nu T^\nu_{\ \mu} = -F_{\mu\gamma} I^\gamma.
  \label{EMC}
\end{equation} 
where 
\begin{equation}
     T^\mu_{\ \nu} = F^{\mu\gamma}F_{\nu\gamma} -
           \frac{1}{4} (F_{\gamma\beta}F^{\gamma\beta})\delta^\mu_{\ \nu}.
  \label{T}
\end{equation}
Since in the Boyer-Lindquist coordinates $\Pd{t} g_{\mu\nu} = \Pd{\phi} g_{\mu\nu} = 0 $, 
the 3+1 splitting of equations (\ref{EMC}) leads to the following energy
\begin{equation}
\Pd{t} e + \vdiv{S} = -(\spr{E}{J}),
  \label{EC}
\end{equation}
and the angular momentum 
\begin{equation}
\Pd{t} l + \vdiv{L} = -(\rho \bE + \vpr{J}{B})\!\cdot\!\bm{m} ,
  \label{AMC}
\end{equation}
conservation laws ($\bm{m}=\Pd{\phi}$). Here
\begin{equation}
  e=-\alpha T^t_{\ t} = \frac{1}{2}(\spr{E}{D}+\spr{B}{H})
  \label{e}
\end{equation}
is called the volume density of energy-at-infinity (or red-shifted energy),
\begin{equation}
  l=\alpha T^t_{\ \phi} = (\vpr{D}{B})\!\cdot\!\bm{m}
  \label{l}
\end{equation}
is the volume density of angular momentum,
\begin{equation}
      \bS=\vpr{E}{H}
  \label{S}
\end{equation}
is the flux of energy-at-infinity energy, and
\begin{equation}
      \bL=-(\spr{E}{m})\bD -(\spr{H}{m})\bB +
              \frac{1}{2}(\spr{E}{D}+\spr{B}{H}) \bm{m}
  \label{L}
\end{equation}
is the flux of angular momentum.

Finally, we remark that $\bB$ and $\bD$ are the magnetic and 
electric fields as measured by local FIDO \cite{K04a}. Thus, according to 
FIDO's observations the energy-density of the electromagnetic field is
$$
\hat{e}= \frac{1}{2}(D^2+B^2)
$$  
and the Poynting flux is
$$ 
      \hat{\bS}=\vpr{D}{B}.
$$
Often the quantities $e,l,\bS,\bL$ are 
described as measured by stationary observers at infinity. This is somewhat 
misleading as in General Relativity only local measurements make sense and this 
is reflected in the notion of local inertial observers. It would be more accurate 
to describe $e,l,\bS,\bL$, as well as $\bE$ and $\bH$, as purely auxiliary fields 
dictated rather by mathematics than by physics. Only at infinity they accept clear 
physical interpretation as locally measured energy, angular momentum etc.

\section{Negative phase velocity waves in the ergoregion of Kerr black hole}
\label{NPV}

Mackay et al.\cite{MLS05b} used eq.\ref{S} in their 
analysis of NPV in black hole magnetospheres. Here we present 
a somewhat different approach which allows us to obtain all the key 
results in such a simple form that we do not need to do any numerical 
calculations. In this approach we utilise the fact that the stress-energy 
momentum tensor of electromagnetic waves in vacuum has the following simple 
form  \cite{LL71}  
\begin{equation}
  T\upin{\nu}\upin{\mu}=\frac{\hat{e}}{\hat{w}^2}k\upin{\nu}k\upin{\mu},
\label{fido-T}
\end{equation}
where $\hat{e}$ is the volume density of the wave energy,
$k\upin{\nu}=(\hat{w},\hat{w}\hat{\bn})$ is the wave vector,
$\hat{w}$ is the angular frequency, and $\hat{\bn}$ is the unit 
vector in the direction of the wave propagation. This representation 
holds for any local inertial observer and our FIDO in particular.     
In the rest of this section we use hat to denote only quantities 
measured by FIDO.  

In the case of vacuum the energy equation (\ref{e}) reads as a proper 
conservation law  
\begin{equation}
  \Pd{t}e+\spr{\nabla}{S}=0,
\label{eq-1.2}
\end{equation}
Notice, that the components of $T^\nu_\mu$ appearing in the 
definition of $e$ and $\bS$ are measured in the coordinate basis of 
space-time coordinates and in general do not represent energy and 
and its flux as measured by FIDO. This becomes true 
only at infinity where the space-time flattens and $t$ runs at the same 
speed as the proper time of FIDO. This remarkable difference 
from the energy conservation law in flat space-time is in fact 
well known and yet sometimes it becomes forgotten \cite{McCall}.

In order to find 
$$
e=-\alpha T^t_t \quad\mbox{and}\quad 
S^i = -\alpha T^i_t
$$ 
we use the transformation between 
the coordinate basis $\{\Pd{t},\Pd{\phi},\Pd{r},\Pd{\theta}\}$ and 
the orthonormal basis of FIDO $\{\ortt{t},\ortt{\phi},\ortt{r},\ortt{\theta}\}$ 
\begin{equation}
\nonumber
  \ortt{t}=\frac{1}{\alpha}\Pd{t}-\frac{\beta^\phi}{\alpha}\Pd{\phi},\quad
  \ortt{\phi}=\frac{1}{\sqrt{\gamma_{\phi\phi}}}\Pd{\phi},\quad 
\end{equation}
\begin{equation}
  \ortt{r}=\frac{1}{\sqrt{\gamma_{rr}}}\Pd{r},\quad
  \ortt{\theta}=\frac{1}{\sqrt{\gamma_{\theta\theta}}}\Pd{\theta}.
\label{eq-1.3}
\end{equation}
Straightforward calculations give us   
\begin{equation}
 \bS = e (\alpha\hat{\bn} -\bbeta) 
\label{eq-s}
\end{equation}
and 
\begin{equation}
 e = \hat{e} (\alpha+\beta n\upin{\phi}), 
\label{eq-e}
\end{equation}
where $\beta=|\bbeta|$ and without any loss of generality we assume 
that the black hole rotates anti-clockwise ($a>0$).
One can see that the poloidal component of the red-shifted energy flux is
\begin{equation}
 \bS_p = \alpha e \hat{\bn}_p
\label{eq-sp}
\end{equation}
and thus $\bS_p$ and $\bn_p$ can have opposite directions (NPV phenomenon) 
provided the energy-at-infinity is negative. This occurs when  
\begin{equation}
 n\upin{\phi} < -\alpha/\beta.
\label{eq-npv}
\end{equation}  
Outside of the ergosphere $\alpha>\beta$ and this condition cannot be  
satisfied. However, inside the ergosphere $\alpha<\beta$ and the red-shifted 
energy is negative provided in the frame of local FIDO the wave propagates in 
the opposite direction to the black hole rotation and its normal is confined 
within the cone with opening angle 
\begin{equation}
  \mu = \arccos(\alpha/\beta).
\label{eq-mu}
\end{equation}
This result, and hence the NPV effect, is not new as it is well known that 
photons can have negative energy-at-infinity inside the black hole ergosphere 
under exactly the same condition as (\ref{eq-npv}). Indeed, if 
$$
  P\upin{\mu}=(\hat{\cal E},\hat{\cal E}\hat{\bn}) 
$$
are the components of the energy-momentum vector of such a particle as 
measured by FIDO then its energy-at-infinity is given by 
$$
 {\cal E}= -P_t = \hat{\cal E}(\alpha+\beta \hat{n}\upin{\phi}). 
$$    

Finally, we remark that only waves created in the ergoregion can have 
negative energy density and they are bound to remain within this region. 
No wave originated outside of the ergosphere will exhibit negative 
phase velocity or the related phenomenon of negative refraction. This 
means that the negative refraction cannot be detected in the images 
of the background sources of radiation.
Since vacuum is a non-dispersive ``medium'' the group velocity of NPV 
waves equals to their phase velocity when measured by FIDO's or 
any other local physical observer. When such waves propagate towards the 
event horizon all such observers will agree that they carry energy and 
information towards the horizon as well.     

\section{Counter-flow of energy}
\label{CFE}

Now consider stationary axisymmetric magnetospheres of black hole filled with highly 
conductive plasma of negligibly small inertia \cite{BZ77,K04a}. 
In such magnetospheres the Lorentz force vanishes and we have 
\begin{equation}
  \rho \bE + \vpr{J}{B} =0,
\label{a3}
\end{equation}
which implies that
\begin{equation}
  \spr{E}{B} = \spr{D}{B} = \spr{E}{J} = 0.
\label{a4}
\end{equation}
Therefore the energy equation (\ref{EC}) becomes 
\begin{equation}
  \spr{\nabla}{S} = 0. 
\label{a5}
\end{equation}
From the stationary version of eq.\ref{Faraday} we find that for axisymmetric fields
$E^\phi=0$ and combining this with eq.\ref{a4} we can write 
\begin{equation}
  \bE = -\vpr{\omega}{B}
\label{a6}
\end{equation}
where $\bmath{\omega} =\Omega \Pd{\phi}$ is a purely azimuthal vector, 
and hence 
\begin{equation}
  \bD = -\frac{1}{\alpha}(\bomega+\bbeta)\!\times\! \bB.
\label{a7}
\end{equation}
Substituting eq.\ref{a6} into the stationary version of eq.\ref{Faraday} we also 
obtain 
\begin{equation}
    \spr{B}{\nabla} \Omega = 0.
\end{equation}
Thus $\Omega$ is constant along $\bB$. It is called  
the angular velocity of magnetic field lines. 
Using the constitutive equations (\ref{E3},\ref{H3}) it is easy to find 
the poloidal component of energy-at-infinity flux 
\begin{equation}
  \bS_p = \vpr{E}{H}= -H_\phi \Omega \bB_p  
\label{a8}
\end{equation}   
and the poloidal component of the energy flux measured by local FIDO 
\begin{equation}
  \hat{\bS_p} = \vpr{D}{B}= -H_\phi (\Omega-\Omega_F) \bB_p
\label{a9}
\end{equation}
where $\Omega_F$ is the angular velocity of FIDO, that is 
$\bbeta=-\Omega_F\Pd{\phi}$ ($\Omega_F$ is positive for an anti-clockwise rotating 
black hole, $a>0$). Combining, these two results we obtain 
\begin{equation}
  \bS_p = \frac{\Omega-\Omega_F}{ \Omega} \hat{\bS}_p.
\label{a10}
\end{equation}
Thus, these two energy flux vectors are parallel if $\Omega>\Omega_F$ 
and anti-parallel if 
\begin{equation}
0<\Omega<\Omega_F.
\label{cond-cf} 
\end{equation}
The {\it counter-flow of 
energy} in the latter case looks similar to the effect of negative phase 
velocity discussed in the previous section. One may understand the direction 
of $\hat{\bS}$ as {\it the direction of electromagnetic field flow, or 
the direction of electromagnetic wind}, and 
the direction of $\bS$ as the direction of energy flow. Then one may 
say that {\it energy is flowing in the direction opposite to that of the field 
flow}. 

Similarly simple calculations allow us to find the following expression for the 
density of energy-at-infinity 
\begin{equation}
  e = \frac{1}{2\alpha}
  \left[ 
   \alpha^2B^2+ B_p^2(\omega^2-\beta^2). 
  \right]
\label{a11}
\end{equation}
Thus, $e$ is negative provided 
\begin{equation}
   \alpha^2B^2+ B_p^2(\omega^2-\beta^2)<0 .
\label{cond-e} 
\end{equation}
This condition is obviously different from (\ref{cond-cf}) and thus in 
contrast to the NPV effect the domains of energy counter-flow and 
negative energy-at-infinity do not coincide in general. In fact, eq.\ref{cond-e} can be 
written as
\begin{equation}
   \Omega^2<\Omega_F^2 - \frac{\alpha^2B^2}{\gamma_{\phi\phi}}. 
\label{a-12}
\end{equation}
Thus, this condition is  stronger than (\ref{cond-cf}) and 
the energy counter-flow can be seen in regions with positive 
energy at infinity.  On the other hand, the condition (\ref{cond-e}) can 
also be written as 
\begin{equation}
  \alpha^2-\beta^2 < -\frac{(\alpha^2 B_a^2 + \omega^2 B_p^2)}{B_p^2}, 
\label{a-13}
\end{equation}
where $B_a$ is the azimuthal component of magnetic field. This 
tells us that negative energy-at-infinity can only be found inside 
the ergosphere, as expected. 

The wind separation surface, $\Omega=\Omega_F$, which is 
called by Okamoto\cite{Oka} ``the effective ergosphere'', 
can be located outside of the real one. 
For example, in the Blandford-Znajek solution for monopole magnetosphere 
$\Omega=\Omega_h/2$ where $\Omega_h=a/2r_+$ is the angular velocity 
of black hole ( $\Omega_h=\lim_{r\to+r_+} \Omega_F$). Thus the 
wind separation surface is defined by 
\begin{equation}
 \Omega_h=2\Omega_F \ntext{or} \Sigma = 8rr_+ 
\label{a-14}
\end{equation}
In the limit $a\to0$ this gives us $r=16^{1/3}$ which higher than 
the radius of ergosphere which in this limit coincides with 
the event horizon, $r_+=2$.     

Additional insight into the phenomenon of the counter-flow of energy 
can be gain from the following simple calculations. Consider a stationary 
monopole magnetosphere of a Kerr black hole. Then in the equatorial plane   
$\bB=\bB_{r}+\bB_{\phi}$ and $\bD=\bD_{\theta}$ (Here indexes are used  
to show directions of vectors.)  Inside the wind separation surface the 
local FIDO measures the ingoing radial Poynting flux 
\begin{equation}
  \hat{\bS}_{r} = \bD_{\theta}\times\bB_\phi 
\end{equation}
which is associated only with components $\bD_{\theta}$ and $\bB_{\phi}$;
$\bB_{r}$ is a passive component and the energy associated with it is 
not transported. The radial component of the energy-at-infinity flux is 
\begin{equation}
  \bS_{r} = \alpha^2\hat{\bS}_r - 
  \alpha \bB_{\phi}\times(\bbeta\times\bB_{r}). 
\end{equation}
It has the opposite direction to $\hat{\bS}_{r}$ provided 
\begin{equation}
  \frac{\beta}{\alpha}|\bB_{r}| > |\bD_{\theta}|.
\label{a-15}
\end{equation}
The volume density of energy-at-infinity is 
\begin{equation}
  e=\frac{\alpha |\bB_{r}|^2}{2} +\frac{1}{2} 
 \left[ \alpha(|\bB_{\phi}|^2 +|\bD_{\theta}|^2) 
  -2 \beta |\bB_{r}||\bD_{\theta}| \right].
\end{equation}
The first term can be understood as the one which is not 
transported by the flow. The second term is negative when 
\begin{equation}
   \frac{\beta}{\alpha}|\bB_{r}| > 
    \frac{|\bD_{\theta}|^2+|\bB_{\phi}|^2}{2|\bD_{\theta}|}.
\end{equation}
Notice that this condition is identical to (\ref{a-15}) if 
like in an electromagnetic wave  $|\bB_{\phi}|=|\bD_{\theta}|$. 
In this case one could conclude that the energy counter-flow occurs 
due to the negative energy-at-infinity of the active component
\begin{equation}
  e_a=\frac{1}{2} 
 \left[ \alpha(|\bB_{\phi}|^2 +|\bD_{\theta}|^2) 
  -2 \beta |\bB_{r}||\bD_{\theta}| \right].
\end{equation}
and write 
\begin{equation}
   \bS_r = \alpha e_a \ortt{r}.
\end{equation}

This analysis shows that the mechanical process of extraction of rotational
energy of black holes by Penrose is a special case of the energy counter-flow 
effect in the sense that the outgoing flow of energy-at-infinity  in the Penrose 
process is inseparable from the negative-energy at infinity of the infalling 
``object''. Although this is true in the case of single particles including  photons
and high frequency electromagnetic waves, in the general case the domains of
negative energy-at-infinity and energy counter-flow are not the same.
Another example if perfect fluid. In this case the stress-energy-momentum
tensor is
\begin{equation}
 T^{\mu\nu}=w u^\mu u^\nu + pg^{\mu\nu},
\end{equation}
where $p$ is the pressure and $w$ is the relativistic enthalpy, and
the condition for energy counter-flow now reads
\begin{equation}
  u_t>0
\end{equation}
whereas the condition for negative energy-at-infinity is weaker
\begin{equation}
  u_t>-p/wu^t.
\end{equation}
Thus, there may exist spatial domains which do not exhibit energy counter-flow 
but where the energy-at-infinity is negative. Only if the fluid is cold, $p=0$,
these conditions become identical. In fact, the stress-energy-momentum of cold
fluid, or dust, has the similar form to that of a high frequency electromagnetic wave,
namely
\begin{equation}
 T^{\mu\nu}=\varrho u^\mu u^\nu,
\end{equation}
where $\varrho$ is the rest mass of fluid and $u^\mu$ is its four-velocity.
In both cases this is just an outer product of lower rank tensors.

\section{Electromotive force of Blandford-Znajek mechanism}

Electromagnetic extraction of rotational energy in the Blandford-Znajek 
process is deeply connected to the electric currents flowing  in black hole magnetosphere. 
Indeed from eq.\ref{a8} if follows that $H_\phi$ must not be zero and 
the stationary axisymmetric version of eq.\ref{Ampere} implies that 
\begin{equation}
 H_\phi(r,\theta) = \frac{I_p(r,\theta)}{2\pi},
\end{equation}
where $I_p(r,\theta)$ is the poloidal electric current flowing inside 
the axisymmetric loop passing through the point $(r,\theta)$. Thus, another 
key question of the Blandford-Znajek process is the origin of the electromotive
force driving these currents. 

In the case of a rotating magnetised star the origin of electromotive force 
is well understood and it is the same as in the case of a unipolar inductor. 
Namely the free electric charges are forced 
to rotate across the magnetic field via collisions with other particles and   
the $\vpr{v}{B}$-force causes the surface charge separation which leads to electric 
field.    
In the membrane paradigm the black hole horizon is 
considered to be analogous to the surface of a rotating magnetised star with 
finite conductivity \cite{TPM}. 
However, this is a purely artificial theoretical construction that does not 
uncover the origin of black hole electric field but hides it instead. 
Moreover, it leads to the perplexing causality issue as explained in 
Punsly \& Coroniti\cite{PCa,PCb}. These authors also pointed out that in 
the Blandford-Znajek solution there was no place for a unipolar inductor. They 
considered this as a crucial weakness of the model and tried to build a theory 
where matter-dominated regions actually develop in black hole magnetospheres 
and play the role of a unipolar inductor. 
However, the development of such matter-dominated regions is not required as 
the electric field in the Blandford-Znajek process is a completely different
purely general relativistic effect.   
It is based on the peculiar electrodynamic properties of vacuum as descried by 
the constitutive equations (\ref{E3},\ref{H3}). 

Suppose a rotating black hole is 
placed in a stationary external field of field $\bB$. 
Then the stationary version of eq.\ref{Faraday} and eq.\ref{E3} yield 
\begin{equation}
 \nabla\times(\alpha\bD) = -\nabla\times(\bbeta\times\bB).
\label{source}
\end{equation}
One can see that the magnetic field and the shift vector combine into a source term for 
the field $\bD$. 
Thus, any local observer will detect not only the magnetic field but also
the electric one. 
It is this electric field that was found in the stationary axisymmetric 
vacuum solution by Wald\cite{Wald}, the result that prompted 
Blandford and Znajek\cite{BZ77} to investigate whether magnetospheres of black 
holes can be similar to pulsar magnetospheres.  

Next issue, first raised in \cite{PCa}, 
is whether this electric field can survive and support poloidal electric currents when 
free charges  (pair plasma) are introduced into the black hole magnetosphere. 
Punsly \& Coroniti\cite{PCa,Pun01} argued that in the absence of a proper unipolar 
inductor the vacuum electric field (\ref{source}) would cause an electric charge separation 
in the magnetosphere that would lead to total screening of this electric field. 
The outcome is a ``dead'' magnetosphere with no poloidal currents and no energy outflow. 
However, a relatively simple analysis shows that this is not true.
Indeed, the total screening would mean 
\begin{equation}
    \spr{D}{B}=0 \ntext{and} B^2-D^2>0.
\label{screen} 
\end{equation}
(If only the first condition is satisfied then there exist locally inertial frames, 
or observers, who see only electric field which clearly indicates incomplete screening.)  
Let us assume that the first condition is satisfied. Then from the constitutive 
equation (\ref{E3}) we have $\spr{E}{B}=0$ and the same analysis as in Sec.5 
leads as to the result (\ref{a7}). This implies that 
\begin{equation}
   (B^2-D^2)\alpha^2 =  
   B^2\left[{\alpha^2}-(\bbeta+\bomega)^2\right] + C_p^2,
\label{B2-D2}
\end{equation}
where 
\begin{equation}
  C_p=(\bomega +\bbeta)\cdot\bB 
\end{equation}    
Using the constitutive equation (\ref{H3}) and the fact that $\bomega$ and $\bbeta$ are 
purely azimuthal we obtain 
\begin{equation}
  C_p=(\Omega-\Omega_F) \frac{I_p}{2\pi}.
\end{equation}    
Thus, in the case of dead magnetosphere $C_p=0$ and we have 
\begin{equation}
   (B^2-D^2)\alpha^2 =  B^2 f(\Omega,r,\theta),
\label{key}  
\end{equation} 
where    
\begin{equation}
f(\Omega,r,\theta)= \alpha^2- (\bbeta+\bomega)^2. 
\label{f}
\end{equation} 
One rather surprising result that follow from this equation is that the strength 
of field $\bB$, which is purely poloidal if $I_P=0$, has no effect on the 
sign of $B^2-D^2$. This simply because a stronger poloidal magnetic field implies 
a stronger electric field due the electromagnetic properties of vacuum 
(see eq.\ref{source}). The only relevant parameter is the angular velocity 
of magnetic field lines $\Omega$. The other important result that follows 
from eq.\ref{key} is that the condition $B^2-D^2 >0$ cannot be 
satisfied everywhere. Indeed, far away from the black hole $a\to 1$, and 
$\bbeta\to 0$ and we obtain 
\begin{equation}
f(\Omega,r,\theta) \to 1- \omega^2 = 1-(\Omega r\sin\theta)^2.
\end{equation}
This requires $\Omega=0$ in order to have $f(\Omega,r,\theta)>0$. However, in this case 
equation (\ref{f}) yields 
\begin{equation}
f(\Omega,r,\theta)= \alpha^2- \beta^2
\label{ergo}
\end{equation} 
which is negative inside the black hole ergosphere. Thus, we conclude that because of 
the existence of ergoregion the two conditions of 
screening cannot be satisfied simultaneously throughout the whole magnetosphere --  
one of them has to give way.

In fact, the condition 
\begin{equation}
  f(\Omega,r,\theta)=0 
\end{equation} 
defines the light surfaces (sometime called the ``light cylinders'', the term 
inherited from the theory of pulsar magnetospheres \cite{GJ69} where the light surface has 
cylindrical shape) of rotating magnetic field lines \cite{K04a}. The world line of 
a point rotating with the angular velocity $\Omega$ is null when 
$f(\Omega,r,\theta)=0$, time-like if $f(\Omega,r,\theta)>0$, and 
space-like if $f(\Omega,r,\theta)<0$. Like in the theory of pulsar magnetospheres 
the superluminal rotation of magnetic field lines outside of the light surfaces 
creates electric field exceeding in strength the poloidal magnetic field \cite{M73,BK00}.  

Thus, the magnetosphere of a black hole cannot be dead provided plenty of charged 
are supplied within a large volume around the hole (The above argument does not exclude 
the possibility of stationary charged clouds or shells  around black holes 
confined to the space between the light surfaces. However, in most astrophysical 
settings the favourable conditions for pair creation are not expected to be confined 
to such a small volume. )
This means that a poloidal current must result from the process of screening of 
vacuum electric field. Indeed, eq.\ref{B2-D2} shows that $I_p\not=0$ is 
a necessary condition for achieving full screening. In fact, in order to sustain such a current 
the electric field should retain small unscreened component along the magnetic field. 
In the singular limit of infinite conductivity this component becomes vanishingly small
and therefore the nature of the electromotive force in the force-free electromagnetic 
models or the ideal MHD models becomes ``camouflaged''.  However, as soon as the condition 
of perfect conductivity is relaxed this ``marginally screened'' electric field 
pops up in the solutions \cite{K04a}. Finally, we point out that in contrast to
poloidal magnetic field the azimuthal magnetic field generated by the poloidal current 
does not strengthen the electric field via eq.\ref{source} and 
this explains how the poloidal current promote screening of electric field.

\section{Discussion}

The analysis given in previous sections leads to an interpretation 
of the Blandford-Znajek process which unites it with all other processes of 
energy extraction from black holes.  

The wind separation surface divides the black hole
magnetosphere into two domains. In the outer domain we find an outgoing 
electromagnetic wind as determined by the direction of the Poynting vector 
measured by the local observer at rest in space. The energy-at-infinity 
in this domain flows in the same direction as the electromagnetic field. 
In the inner domain the electromagnetic wind blows towards the black hole 
and in the opposite direction to the flow of energy-at-infinity. One may compare  
the outgoing wind with the outgoing particle of positive energy-at-infinity in the Penrose 
process and the ingoing wind with the ingoing particle with negative energy-at-infinity.     
The paired-wind structure of flows in black hole magnetospheres has 
been pointed already in Blandford-Znajek\cite{BZ77} and later in the MHD models, e.g. 
\cite{TNTT,Oka92,BK00}. However, the flow direction was discussed in terms of the 
particle flux which is not particularly relevant given the very small 
contribution of particles to the total energy transport (which is dominated by the 
electromagnetic field). 

It appears that the mechanical process of extraction of rotational 
energy of black holes by Penrose is based on a somewhat 
special case of the phenomenon which we call ``the energy counter flow''. 
This case is special in the sense that the energy extraction is 
inseparably linked to the negative-energy at infinity of the infalling ``objects''. 
Although this is true in the case of  single particles including  photons,  
high frequency electromagnetic waves, and flow of dust, 
in the general case the domains of 
negative energy-at-infinity and energy counter-flow are not the same. 
Moreover, as we have seen in Sec.5, the wind separation surface 
can be located outside of the ergosphere (in agreement with 
Okamoto\cite{Oka92,Oka}, whereas the negative energy particles of Penrose 
process can only be created inside the ergosphere.  

The ergosphere, however, plays a key role in the Blandford-Znajek process as well. 
This becomes clear when we comprehend the nature of the electromotive force that 
drives the poloidal currents of black hole magnetosphere and causes its  separation 
into ingoing and outgoing wind domains. Essentially, what drives the currents is 
the vacuum electric field electric field generated in the vicinity of rotating 
black hole placed into magnetic field. As we have seen in Sec.6 it is the existence 
of the ergoregion that makes impossible the total screening of vacuum electric field 
via electric charge separation alone. 
Rather unusually, the poloidal electric current itself 
becomes an integral part of the screening process which is never fully complete 
and leaves small unscreened component of electric field that is required to support 
this poloidal current.  As the result, and in great contrast to the magnetospheres 
of neutron stars and other less extreme astrophysical objects, there is no need in 
the traditional ``unipolar inductor'', that is a matter-dominated magnetised 
rotating object. One may say that it has been replaced by the rotating space of 
Kerr black hole (This is what is meant when the inertial frame dragging 
effect, which manifests itself via the shift vector $\bbeta$, is described as the 
origin of the black hole electromotive force \cite{BK00,L06}.)   

The notion of ``horizon membrane'' endowed with the properties of unipolar inductor \cite{TPM} 
served the purpose of explaining the electrodynamics of black holes in terms of 
classical physics and facilitating its application by wide astrophysical community, not 
intimately familiar with the intricacies of General Relativity. 
The numerous applications of this approach by astrophysicists prove its merits. 
However, this construction is purely artificial and introduces false knowledge which, 
when taken too seriously, can lead to wrong conclusions and controversy\cite{PCa,Pun01,BK00,K04a}. 
The analysis presented in Sec.3 
shows an alternative approach which, like the membrane paradigm, allows us to use 
the language of classical physics in black hole electrodynamics. Indeed, 
the 3+1 equations of black hole electrodynamics can be written in exactly the 
same form as the traditional Maxwell equations in matter and hence  
the  vacuum around black holes can be treated as an electromagnetically active medium.   
The clear advantage of 
this approach, which was first suggested by Tamm at the beginning of the last 
century\cite{Tamm}, is that it does not introduce any fictional elements and is 
fully based on rigorous mathematics. As the result, and in contrast to 
the Membrane paradigm, it does not need to involve a blind faith in the validity 
of its applications.             

Another issue related to the nature of black hole electromotive force 
is the so-called ``Meissner effect of black hole
electrodynamics,'' that is the expulsion of magnetic flux from the
horizon of rapidly rotating black holes in axisymmetric vacuum solutions. 
Within the membrane paradigm this suggests a reduction in the efficiency of the
Blandford-Znajek mechanism as the magnetic field becomes detached from 
the ``unipolar inductor'' \cite{BJ} . However, once we have realised that it is 
not the horizon but the ergoregion that is important for the mechanism this 
issue goes away. Moreover,  the Meissner effect is not found in conductive 
magnetospheres \cite{KM07}.

The creation of charged particles in black hole magnetospheres is another 
important element of the Blandford-Znajek process.  It has been argued that
particles are created in potential gaps developed in black hole 
magnetospheres, either at the wind-separation surface\cite{TNTT,Oka} or at the surface 
where the density of Goldreich-Julian charge changes its sign \cite{BIP}, 
and that the micro-physical processes in the gaps determine the structure of the 
magnetosphere. For example, Takahashi et al.\cite{TNTT} make a rather vague comment 
on the possibility of Penrose process taking place at the wind separation surface 
and call the region inside the surface the region of negative energy. However, closer 
look reveals that what they actually mean by this is the opposite directions of the 
particle and the total energy-at-infinity flows, that is another version of the energy 
counter flow effect. Okamoto\cite{Oka} goes further and concludes that not just 
one but two unipolar inductors exist at both sides of the wind separation surface and 
support two electric circuits, the inner and the outer one, with two antiparallel currents 
of almost equal strength flowing across the magnetic field at this surface. 
It is hard to see anything in this model but an artificial construction, similar to 
the horizon membrane\cite{TPM}, whose whole purpose is to 
make the black hole magnetospheres look more like the magnetospheres of normal stars. 
A proper unipolar inductor would surely be a massive matter dominated rotating object 
which cannot develop as the only way of matter creation at the wind separation surface
is the pair cascade in strong electromagnetic field.  

The electric field in the gap has to be strong enough to accelerate electrons and 
positrons to the energies sufficient for initiating the pair cascade and, strictly 
speaking, the force-free approximation breaks down inside the gap. 
If the required potential 
drop is not small compared to the global potential drop in the magnetosphere then 
the gap width can be noticeable and the magnetic field lines can rotate with 
noticeably different angular velocities on both sides of the gap. However, 
this does not seem to be the case in most astrophysical settings as it requires either 
very slow rotation of the black hole or very weak magnetic field.      
Another important issue is whether the gaps can develop a steady-state structure 
which passively adjusts to the global current structure of the force-free magnetosphere, 
transmitting the required amount of electric current. If not then there can be no 
steady-state solution for the global magnetosphere too and one can expect temporary 
sparks of particle creation in various locations where the screening of vacuum 
electric field becomes ineffective, due to the lack of electric charges, or force-free waves 
radiated from the gap with time-variable structure \cite{L05}.  
This problem requires further investigation. In fact, pair creation does 
not have to be confined to potential gaps in magnetosphere and in various astrophysical 
settings it may occur over much larger volumes. If, for example, the accretion disk can provide 
sufficient flux of photons with energy above the rest-mass energy of electron-positron 
pair then the pairs can be produced via the two-photon reaction\cite{Phinney}. 
In the case of Gamma Ray Bursters the pairs are readily produced
via the neutrino-antineutrino annihilation \cite{Eichler,MW99}

\begin{acknowledgments} 
The author would like to thank Prof.V.Beskin for many helpful 
discussions. 
\end{acknowledgments}


\end{document}